\documentclass[aps,prl,reprint,groupedaddress]{revtex4-1}

\usepackage{graphicx}
\usepackage[T1]{fontenc}
\usepackage{mathptmx}
\usepackage{amsmath}
\usepackage{siunitx}

\begin{document}


\title{Surface Crystallization of Monoatomic Pd Metallic Glasses}


\author{Y. Huang}

\author{L. Xie}

\author{D.S. He}

\email{Corresponding author.\\heds@sustech.edu.cn}

\author{J.Q. He}
\email{Corresponding author.\\he.jq@sustech.edu.cn}

\affiliation{Department of Physics and Pico Center, Southern University of Science and Technology, Shenzhen 518055, China}

\date{\today}

\begin{abstract}
Crystallization from an amorphous atomic structure is usually seen as a spontaneous process in pursuit of a lower energy state, but for alloy systems it is often hard to elucidate because of the intrinsic structural and compositional complexity. Here, by means of electron beam irradiation, we found surface-limited, and thus size-dependent crystallization in a system of monoatomic Pd metallic glass, which is ascribed to the structural differences between the surface and the interior. The equilibrium thickness of the surface crystallization is controllable, presenting a promising approach to fabricate novel nanostructures. The investigation is believed to provide a general understanding of solid amorphous-to-crystalline phase transition from the nanoscale to the bulk size.
\end{abstract}


\maketitle

Phase transition in solids is very common in nature for both organic and inorganic materials \cite{sharma2015developments,xu2018boosting,wu2014origin,umair2019novel}, and it is made advantage to write information \cite{wuttig2007phase,raoux2010phase}, store thermal energy \cite{sharma2015developments,pielichowska2014phase} and design high performance thermoelectric materials \cite{wu2015advanced,chang20183d}. Understanding the phase transition in amorphous materials is extremely challenging because of the intrinsic structural complexity, e.g., multiple structural motifs in metallic glasses \cite{cheng2009atomic}. Among the phase transitions of simple amorphous materials, the amorphous-to-crystalline transition is probably the easiest to observe, identify, and understand. It can be seen as a process of proceeding to a lower free energy state ($\Delta G\leq 0 $). The driving factor and kinetic pathway of this process, especially at the reduced scale, is important to the knowledge of stabilizing and designing complex crystallographic phase of nanomaterials. However, most amorphous materials discovered to date contain multiple elements, that exhibit spatially compositional heterogeneity \cite{zhu2017correlation}. The compositional complexity, together with the structural complexity in the atom disorderliness, creates enormous local energy minima on the pathway of the amorphous-to-crystalline transition, making reliable experimental investigation, and thus true mechanistic elucidation, at or lower than the nanoscale (for example, at the surface or interface) almost impossible. In this sense, single element metallic glass, which is not only for its great potential applications such as catalysis \cite{yang2018amorphous} and gas purifying \cite{DongshengHe2020}, but also for the ability of generating good contrast in the electron microscopes, is the ideal benchmark for the mechanistic study of the amorphous-to-crystalline transition. So far, occasional publications report that electron beam can cause complete solid crystallization in the nanosized monoatomic metallic glass \cite{zhong2014formation,tang2015amorphization}, but systematic  studies were failed to be provided.  As a consequence, the true mechanism for the amorphous-to-crystalline transition, i.e., the intrinsic metastability or the size-dependent instability of the amorphous phase, remains unknown.

In this letter, we take the monoatomic Pd metallic glass nanoparticles as the model system. The surface dynamics of nanosized monoatomic Pd metallic glass were investigated under the irradiation of an intensive electron beam. We observed the disorder-order transformation process on the surface of nanosized Pd metallic glass and explored the size dependence and temperature effect of this system. This phenomenon can be explained by the free energy model and we highlight that the structural differences on the surface of metallic glass play an important role in surface crystallization.

\begin{figure}
	\centering
	\includegraphics[scale=0.2]{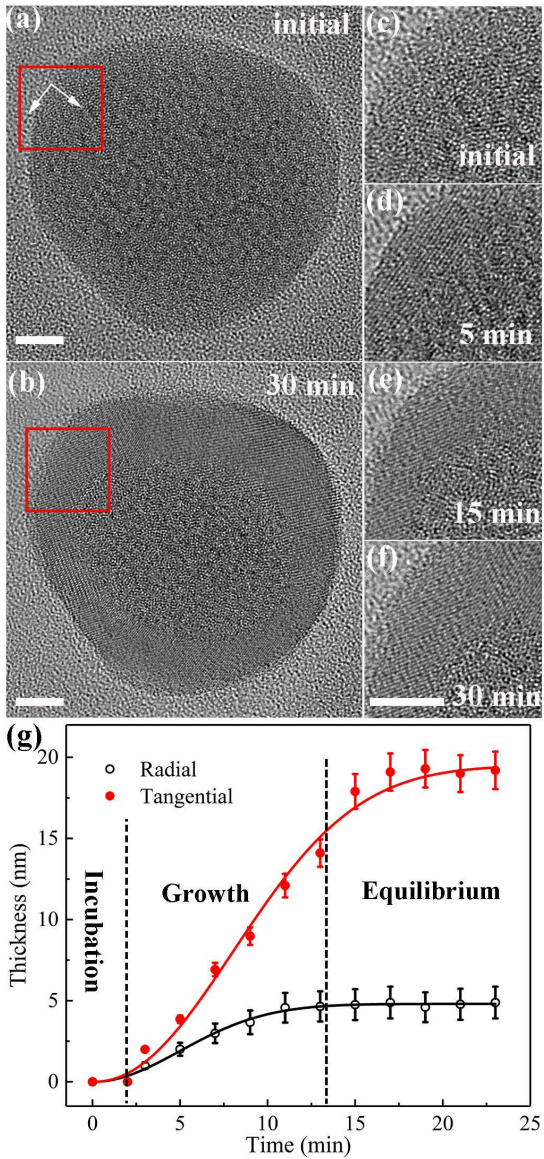}
	\caption{Transformation kinetics of monoatomic Pd glassy nanoparticle with $ d\sim 30 $ nm irradiated by electron beam. (a,b) HRTEM images of the initial and equilibrium nanoparticle, respectively (scale bar: 5 nm). The square and arrows in (a) mark the first nucleus site and the direction of grain growth, respectively. (c)-(f) Magnified HRTEM images of the region marked with a red square in (a) and (b), showing the formation and evolution of crystalline layer of amorphous Pd nanoparticle at different times (scale bar: 5 nm).  (g) The transformation kinetic curve of the particle. The solid line is the fit using the modified JMA equation.} 
\end{figure}
In our experiment, the amorphous single-element Pd nanoparticles were prepared using a simple heat treatment as described by our group \cite{DongshengHe2020}. These nanoparticles were exposed to an intensive electron beam in transmission electron microscope (TEM) and a CCD underneath was used for imaging simultaneously. By doing this way, it enables us direct observation of the structural transformation under strong electron irradiation. The fluorescence screen, which has been pre-calibrated with the Faraday cup, was used to estimate the dose rate. In a typical experiment with a dose rate of $6680\ e/ \si{\angstrom} ^2s$ at 300 keV energy, the structural change is shown in Fig. 1 and Movie S1. Figs. 1 (a) and (b) display the initial and equilibrium state of the amorphous Pd nanoparticle, respectively. Fig. 1(a) shows that the nanoparticle located on the silicon nitride (SiNx) substrate presents a maze-like contrast, indicating a fully amorphous nature. After prolonged irradiation to 30 min, as shown in Fig. 1(b), a poly-crystalline layer formed at the surface of the nanoparticle, while the core of the nanoparticle still remained disordered. Several snapshots of the red square region in Figs. 1(a) and (b) from Movie S1 are displayed in Figs. 1(c)-1(f), which present the time-elapsed transformation process of the first crystalline grain. After 5 min of irradiation, one-dimensional lattice contrast appears on the left-hand corner of the particle, indicating that the amorphous atom arrangement turns into small crystal. Once the crystal nucleus has formed, the thickness of the crystalline grain increases fast with the prolonged irradiation time. The size change of the crystalline grain is plotted along the tangential and radial directions as a function of irradiation time. As shown in Fig. 1(g), the surface crystallization experiences three stages: incubation, growth, and equilibrium. The incubation stage was observed to be lasting for about 3 - 5 min. Immediately after the nuclei come into being, the crystallization spread around the nanoparticle until it reached an equilibrium state. The growth rate of the tangential direction is measured faster than that in the radial side, but it reaches to equilibrium state more slowly than that in the radial direction. This time-dependent phase transformation process was fitted using modified phenomenological Johnson-Mehl-Avrami (JMA) model \cite{avrami1941granulation,yinnon1983applications}, given by $ f(t)=A \left[ 1-exp(-kt^n)\right] $, where $ A $, $ k $, $ t $, and $ n $ denote the shape factor, kinetic constant, time, and Avrami exponent, respectively. By fitting the data (solid lines in Fig. 1(g)), we obtained $n \sim 2$. In addition, the crystalline layer appears with polycrystals because of the multi-nucleus sites near the surface (see Movie S1).

 \begin{figure}
	\centering
	\includegraphics[scale=0.2]{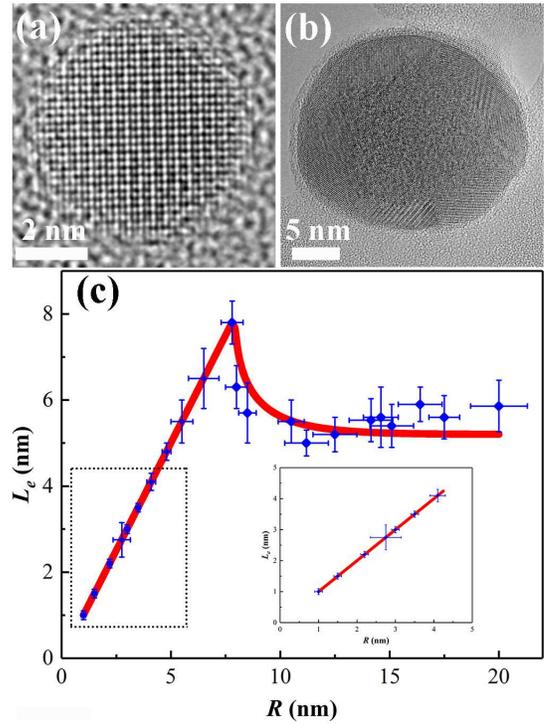}
	\caption{(a)-(b) HRTEM images of the amorphous-to-crystalline transition with different particle sizes. (c) Equilibrium thickness of surface crystallization as a function of particle radius. The solid line is the fit to the data and the inset shows the magnified curve of the square box.}
\end{figure}
Through the above process, it is noticeable that the electron-beam-induced crystallization is only limited at the surface, with an equilibrium thickness ($L_e$) of about 5 nm, which implies a size-dependent effect. Thus, we have investigated 22 nanoparticles with radius $ R $ from 2.5 nm to 20 nm in the same way and the experimental results are displayed in Fig. 2. It was found that amorphous nanoparticles with a radius of less than 7 nm were completely crystallized into single crystals under the intensive electron beam exposure (Fig. 2(a)). For nanoparticles with a radius of more than 7 nm, they tend to form amorphous-crystalline core-shell structures with equilibrium thickness of 5 nm (Fig. 2(b)). Fig. 2(c) summarized the $L_e$ as a function of \textit{R} and the mechanism will be discussed in the later context. 

In addition, the threshold dose rate to trigger the crystallization was measured as a function of incident beam energy. The threshold dose rate is defined in such a way that can trigger surface crystallization for nanoparticles with diameter $d \sim 25 - 35$ nm within 10 min. The result is presented in Fig. 3(a). Even with accelerating voltage as low as 80 keV, the electron beam can still induce surface crystallization with the same equilibrium thickness. The threshold dose rate was found to slightly decrease with the incident beam energy. According to the energy transfer equation $E_{max}=2U_p(U_p+2mc^2)/Mc^2$ \cite{williams2009carter}(where $U_p$, $M$, $m$, and $c$ denote the kinetic energy, nucleus of mass, electron of mass, and velocity of light, respectively). The calculated maximum energies $E_{max}$ received by a target atom under 300 keV and 80 keV are approximately 8 eV and 1.8 eV, respectively, which are far less than the experimentally reported displacement energy of Pd atom ($34\pm2$ eV \cite{jimenez1967radiation}). According to the previous research \cite{iijima1986structural,wang2012determination}, the momentum transfer between the fast electrons and target atoms, alters the atom arrangements, and thus causes crystallization. It shows the extreme vulnerability of the surface structure of the metallic glass.

\begin{figure}
	\centering
	\includegraphics[scale=0.24]{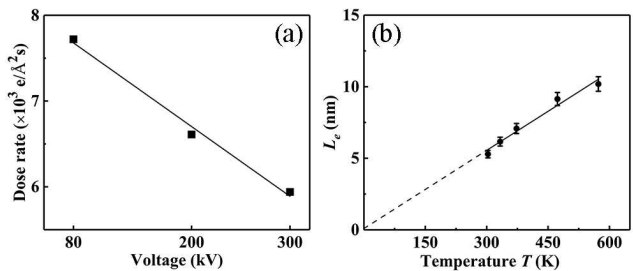}
	\caption{(a) Relationship between the threshold dose rate required to trigger the surface crystallization and the voltage. (b) The equilibrium thickness regulation as a function of temperature. The solid line is the fit with an extrapolation to 0 K (dashed line).}
\end{figure}
We also investigated the temperature effect on the surface crystallization. We were able to apply heating to the nanoparticles due to the unique preparation method we used \cite{DongshengHe2020}. We found that no crystallization occurred even at a temperature of 400 K higher than room temperature (RT) when the electron dose rate was lower than the threshold. It reflects that the surface crystallization can not be triggered by temperature alone. However, our experiment found that temperature, $T$, can play a unique role to modulate $L_e$. The nanoparticles were designed to be annealed at $ T $ = 323 K, 373 K, 473 K, and 573 K, and exposed to electron beam with a dose rate above the threshold shown in Fig. 3(a) to measure the temperature effect on  the equilibrium thickness of surface crystallization. Fig. 3(b) shows the evolution of equilibrium thickness as a function of $T$. $L_e$ is found linearly increased with $T$. This provides a way to modulate the equilibrium crystalline thickness from $L_e\sim$ 5 nm at RT up to $L_e\sim$ 10 nm at $T=$ 573 K. If extrapolating the linear fitting curve of tailored equilibrium crystalline thickness to absolute temperature (0 K), the intersection of curve and coordinate is zero. That implies the surface crystallization can be avoided by freezing the movement of  atoms.

The experimental fact that the crystallization is surface-limited is an indication of a difference between the surface and the interior of the nanoparticles. The lower activation energy at the surface explains the initialization of the crystallization \cite{stevenson2008surface,cao2015high}, but what is the stopping factor for the crystallization? Interfacial energy is not a major factor prohibiting further crystallization because in our case the area of the interface decreases as the crystallization proceeds inwards. Thus, nanobeam electron diffraction (NBD) technique \cite{huang2008coordination,pennycook2011scanning} was performed to study the surface structure of the amorphous Pd nanoparticle. As shown in Fig. 4(a), a probe with the size of 5 nm was used to scan the nanosized particle from particle surface to interior with the step of 0.628 nm; 46 patterns were acquired (Fig. S2). A typical NBD pattern was illustrated in Fig. 4(b), where sharp discontinuous speckles were clearly seen at \textit{\textbf{k}} $\sim$ 4.5 1/nm, signifying the short-rang-order (SRO) of the metallic glasses. The background-subtracted rotational average intensity profiles of the whole 46 patterns were extracted and presented in Fig. 4(c). For the first group of peaks (3.5 1/nm < \textit{\textbf{k}} < 6 1/nm) excited from the sample (red and blue curves in Fig. 4(c)), obvious asymmetrical shapes and the shape evolution were observed as the probe moves toward interior, indicating structural differences of the nanoparticle \cite{mauro2014structural}. To further analyze these profiles, the integrated intensity (\textit{\textbf{I}}), the location of intensity maximum (\textit{\textbf{K}}), and ratio between left half peak and  right half peak (see Fig. S3) of the first group peaks as a function of distance are plotted in Fig. 4(d). The integrated intensity was used to identify the edge of the nanoparticle, where it exhibits a sharp increase (marked by a red dashed line in Fig 4 (d)). The locations of the intensity maximum gradually decrease and finally reach a constant when the probe moves from the surface to the interior. According to the reciprocal theorem, it represents the most probable atom-atom distance  experiences contraction at the surface. However, this contraction is relived by less probable atom bonding, i.e. the left side of the peak is wider than the right side. Both of the location of the intensity maximum and the ratio have the same decaying trend and length (region between the blue dashed line and the red dashed line in Fig. 4(d)), which indicates that they are correlated, as a result of  the structural evolution. The decaying length is $ \sim $ 13 nm,  slightly larger than the sum of the probe size ($ \sim $ 5 nm) and equilibrium thickness $ L_e $ at this size ( $ \sim $ 5 nm), implying a energy barrier is needed to be overcome for crystallization, which will be discussed in the following context. 

\begin{figure}
	\centering
	\includegraphics[scale=0.2]{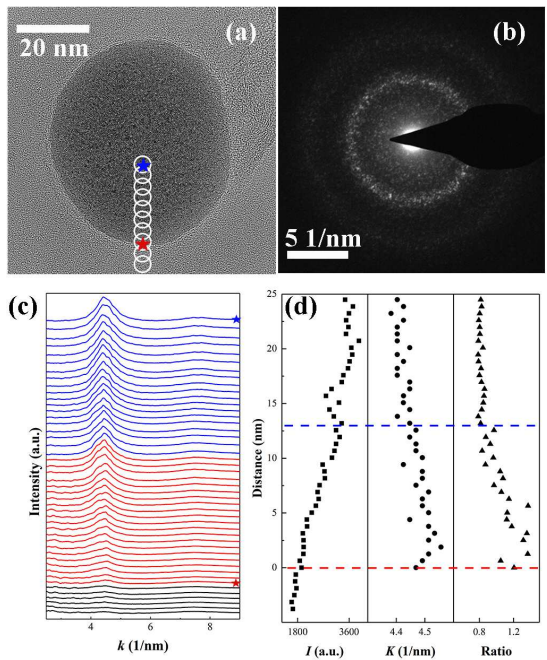}
	\caption{NBD patterns of the Pd monoatomic metallic glass acquired in a line scan mode. (a) The HRTEM image and the scheme of line scan. The red and blue stars denote the onset of surface and the endpoint of acquiring, respectively. (b) A typical NBD pattern. (c) The series of background-subtracted rotational average intensity profiles extracted from NBD patterns. (d) The profile of integrated intensity (\textit{\textbf{I}}), the location of intensity maximum (\textit{\textbf{K}}), and ratio between left half peak and right half peak of the first group peaks as a function of distance.}
\end{figure}
The arrangement complexity of atoms near the surface enables us to model the Gibbs free energy of the amorphous Pd nanoparticles as a gradually decaying function from the surface to interior. Consequently, the free energy change $\Delta G$ of the amorphous-to-crystalline phase transformation can be depicted based on a sphere model, which assuming that the crystallization occurs in the entire surface and grows uniformly toward interior:
	\begin{equation*}
	\begin{aligned}
	\Delta G = &\int_{0}^{L_e} (G_{ac/v} - G_{a/v})\, dv+4\pi (R-L_{e})^2\sigma_{int}\\
	                 &+4\pi R^2\sigma_{c}-4\pi R^2\sigma_{a}
	\end{aligned}
	\end{equation*}
where $ G_{ac/v} $, $ G_{a/v} $, $ R $, $ \sigma_{int} $, $ \sigma_{a} $, and $ \sigma_{c} $ denote the density of free energy of amorphous/crystalline core/shell structure and the initial nanoparticle, the size of the nanoparticle, the interfacial energy of the boundary of core/shell structure, the surface energy of the pristine amorphous state, and induced crystal layer, respectively. To understand the stopping factor of crystallization, taking the derivative $ \partial \Delta G/\partial L_{e}=0 $, it can be seen that the surface energy terms $4\pi R^2 \sigma_c$ and $4\pi R^2 \sigma_a$ do not affect the equilibrium thickness of the surface crystallization because they are independent of $L_e$. Assuming a symmetric exponential decay of the $G_{a/v}$ (see SI for detail), we are able to obtain the relationship between the size of the nanoparticles and the equilibrium thickness of the surface crystallization. By fitting the data (see the solid line in Fig. 2(c)), it shows that, rather the  interfacial energy, it is the lowering of Gibbs free energy of the interior part of the nanoparticle stops further crystallization. In addition, it also explains the size-dependence of the crystallization thickness, where the complete crystallization of small nanoparticles ($ R \leq 7 $ nm) is caused by the overlapping of the tail of the decaying Gibbs free energy of the amorphous nanoparticle. This model also predicts that the equilibrium thickness will remain about 5 nm if we further increase the size of the nanoparticles.

In conclusion, we have reported a novel surface crystallization phenomenon with nanosized Pd metallic glass in response to intensive electron beam. This is caused by the structural differences on the surface. The equilibrium thickness of the surface crystallization is size-dependent and controllable by heating. The results present a novel approach to understanding the solid amorphous-crystal transition and further engineering the crystallographic phase of the nanoparticles.

The work at SUSTech was supported by the National Natural Science Foundation of China (Grant No. 11934007, 11874194, 51632005, and 51602143), the leading talents of Guangdong Province Program (Grant No. 00201517), and the Science and Technology Innovation Committee Foundation of Shenzhen (Grant No. KQTD2016022619565991 and ZDSYS20141118160434515). The authors also acknowledge the support of Pico Center and SUSTech Core Lab Facilities.


\end{document}